# Emulation of Synaptic Plasticity on Cobalt based Synaptic Transistor for Neuromorphic Computing

*P. Monalisha [1], P. S. Anil Kumar [1], X. Renshaw Wang [2,3], S. N. Piramanayagam [2]*

[1]*Department of Physics, Indian Institute of Science, Bangalore, India, 560012*

[2]*Division of Physics and Applied Physics, School of Physical and Mathematical Sciences, Nanyang Technological University, 637371, Singapore*

[3]*School of Electrical and Electronic Engineering, Nanyang Technological University, 637371, Singapore*

**Abstract:**

Neuromorphic Computing (NC), which emulates neural activities of the human brain, is considered for low-power implementation of artificial intelligence. Towards realizing NC, fabrication, and investigations of hardware elements such as synaptic devices and neurons are essential. Electrolyte gating has been widely used for conductance modulation by massive carrier injections and has proven to be an effective way of emulating biological synapses. Synaptic devices, in the form of synaptic transistors, have been studied using a wide variety of materials. However, studies on metallic channel based synaptic transistors remain vastly unexplored. Here, we have demonstrated a three-terminal cobalt-based synaptic transistor to emulate biological synapse. We realized gating controlled multilevel, nonvolatile conducting states in the proposed device. The device could successfully emulate essential synaptic functions demonstrating short-term and long-term plasticity. A transition from short-term memory to long-term memory has been realized by tuning gate pulse amplitude and duration. The crucial cognitive behavior viz., learning, forgetting, and relearning, has been emulated, showing resemblance to the human brain. Along with learning and memory, the device showed dynamic filtering behavior. These results provide an insight into the design of metallic channel based synaptic transistors for neuromorphic computing.

**Keywords:** Synaptic transistor, Metallic channel, multilevel states, synaptic plasticity, ionic-liquid gating, Neuromorphic computing

**Introduction:**

The human brain can solve complex problems with ultralow energy consumption of $\approx$ 20 W.[1, 2] In the human brain, the densely packed neuronal network consists of ~$10^{11}$ neurons interconnected by ~ $10^{15}$ synapses, where signal transmission occurs across the synapses. Whereas the existing digital computers are implemented on von Neumann architecture, which faces the limitation of energy inefficiency and the challenge of the von Newman bottleneck.[3]Therefore, researchers are looking for hardware implementation of brain-inspired computing that physically achieves massive parallelism, ultralow power consumption, self-learning, and fault tolerance. In this connection, the development of electronic devices mimicking synaptic and neuron functions is crucial. Artificial synapses have been developed based on two-terminal devices such as phase-change memory [4, 5], atomic switch [6, 7], memristor [8–10][11], and have emulated the essential synaptic functions. The two-terminal domain wall magnetic tunnel junction (MTJ) devices based on spin-transfer torque are also potential candidates for synapses.[12, 13] However, in two-terminal devices, the learning is usually achieved by feedback from the post neuron to the synaptic device. Moreover, during the learning process, the signal transmission is highly depressed.[14, 15] Therefore, complete emulation of the synapse is limited as learning and signal transmission could not be realized simultaneously. However, more recently, a three-terminal synaptic device was proposed by electrolyte gating. In this device, signal transmission occurs in the channel, while learning is achieved independently using the gate terminal, leading to the complete emulation of a synapse. Various materials such as organic materials [16–18][19], metal oxides [20–22][23], 2D materials [24–26] have been used to emulate the synaptic functions.

Despite remarkable progress in electrolyte-gated synaptic transistors using different channel materials, the study of metallic channel based synaptic transistors remains massively unexplored. Several reports illustrate the modification of the electrical properties of metallic thin films utilizing electrolyte gating due to massive carrier injection.[27–29] In this regard, metallic cobalt thin films are of substantial interest in academic and industrial applications due to their attractive physical, mechanical, electrical, and magnetic properties.[30–32] [33] Cobalt thin films are used in a wide variety of technological applications: including integrated circuitry (IC) devices [34][35], magnetic information storage [36], and sensor systems [37]. Furthermore, sthe spin degree of freedom of cobalt can be tuned by electrolyte gating to tailor these synapses for different applications in the future. These attractive features of cobalt thin film offer the chance to design three-terminal synaptic devices for neuromorphic computing applications.

In this work, we have demonstrated a three-terminal cobalt-based synaptic transistor by ionic liquid gating to emulate a biological synapse. This is the first-ever demonstration of a synaptic transistor using a metallic channel. We have successfully imitated essential synaptic functions including excitatory/inhibitory postsynaptic conductance (E/IPSC), paired-pulse facilitation/depression (PPF/D), long-term potentiation/depression (LTP/D) in the proposed device. The transition from short-term memory (STM) to long-term memory (LTM) has been realized by tuning the gate pulse amplitude and duration. Moreover, the essential psychological behaviour of learning, forgetting, and relearning of the human brain has been mimicked. Furthermore, high/low pass filtering behaviour of the synaptic device has been demonstrated. The metallic channel-based artificial synapse may find potential applications in hardware neural networks for neuromorphic computing.

**Results and Discussion:**

For the three-terminal cobalt-based synaptic transistor, the channel consists of Ta (3 nm)/Co (10 nm)/Ta (2 nm) stack on a thermally oxidized silicon substrate. The bottom Ta layer was used as a buffer layer between the substrate and Co to increase the adhesion. Whereas the top Ta layer was used as a capping layer to protect cobalt from being oxidized by the ambient. The channel was fabricated using standard photolithography and a subsequent lift-off process. The coplanar gate electrode consists of Ta (5 nm)/Cu (90 nm)/Ta (5 nm), was patterned using photolithography, and deposited via sputtering. The further details of device fabrication can be found in the experimental details. The current-voltage relationship of the device showed linear behavior, ensuring good ohmic contacts under the electrodes (Figure S1b, supplementary information). As received IL, 1-Ethyl-3-methylimidazolium bis(trifluoromethylsulfonyl)imide (EMIM-TFSI) was used as the gate dielectric for the present study. The IL has a high capacitance value of 2.5 µF/cm$^2$ at 70 Hz, as evident from the $C\sim f$ measurement, shown in supplementary (Figure S1a). A small drop of IL was dropped on the device covering the channel and a larger part of the gate electrode. The schematic diagram of the three-terminal cobalt-based synaptic transistor (with electrical connection) and the channel multilayer stack

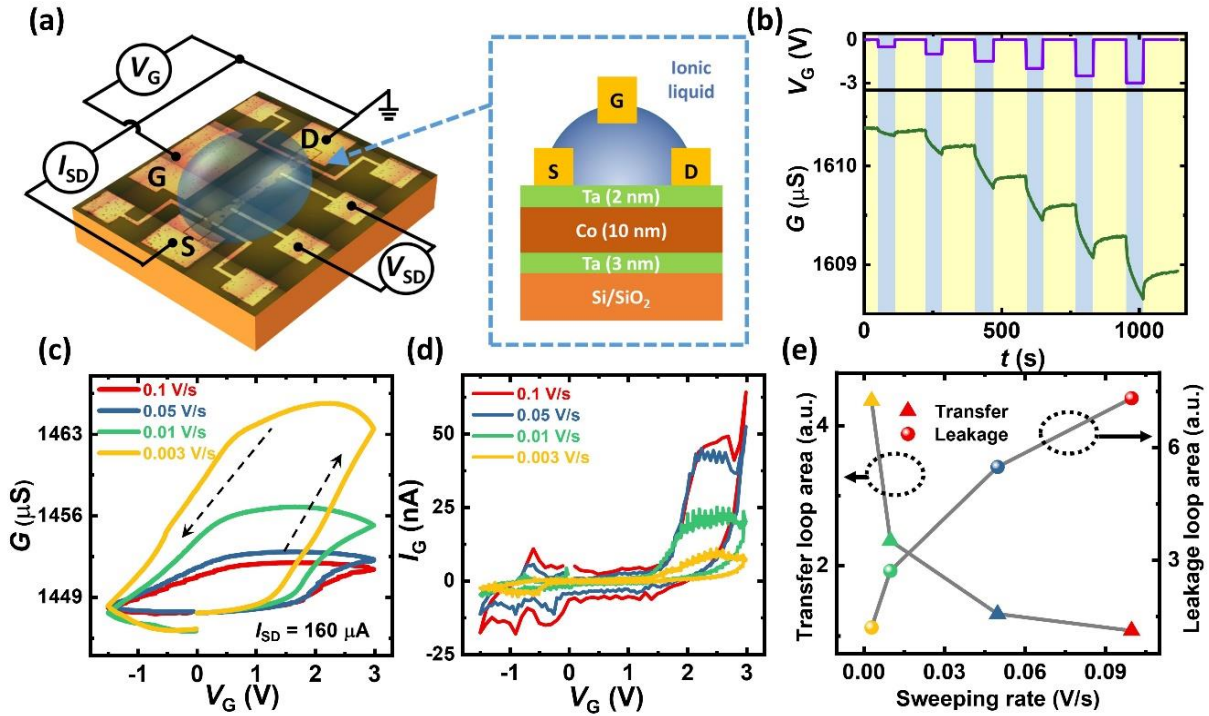

**Figure 1** (a) Schematic diagram of the three terminal cobalt based synaptic transistor. (b) Multilevel nonvolatile conducting states realized by applying series of negative gate voltage pulses of different amplitude (in sequence of $V_G$ = -0.5, -1, -1.5, -2 and -2.5 V) and of same duration ($t_p$ = 60 s), spaced apart by 120 s. $I_{SD}$ = 160 µA. c) Transfer curves measured at different gate voltage sweeping rates (0.1, 0.05, 0.01 and 0.003 V/s), $I_{SD}$ = 160 µA. (d) The corresponding leakage current at different sweeping rates. (e) Area under the transfer curve loop (left) and area under the leakage current loop (right) - as function of gate voltage sweeping rate.

are shown in Figure 1a. A small direct current ($I_{SD}$=160 µA) was applied across the source (S) and drain (D) electrode and the channel conductance was monitored. The gate voltage ($V_G$) was applied directly on the gate (G) electrode, and the drain terminal was grounded throughout the measurement. All the electrical measurements were carried out in a vacuum of $10^{-3}$ mbar to avoid ambient humidity.

For the electrolyte-gated cobalt-based synaptic transistor, the transfer curve was measured by sweeping the gate voltage from 0 to 3 V, then from 3 to -1.5 V, and finally back to 0 V, and monitoring the channel conductance as shown in Figure 1c. A giant anticlockwise hysteric transfer curve indicates the nonvolatile and reversible change of channel conductance, suitable for emulation of synaptic properties. The Leakage current $I_G$ was monitored simultaneously, shown in Figure 1d. As the IL is ionically conducting and electrically insulating, $I_G$ was negligible. $I_G$ (~ nA) was orders of magnitude smaller than input current (~ 160 µA), indicating that conductance modulation was unaffected by the leakage current.

The study of IL gating is widespread for dramatic conductance modulation by massive carrier injection.[29, 38] In IL gated cobalt-based synaptic transistor, the channel conductance was increased (decreased) during positive (negative) gating due to the formation of an electric double layer (EDL) at the channel/IL interface. Additionally, the IL contains traces of water, which undergoes hydrolysis splitting and releases protons and hydroxyls ions above a threshold ($V_T$).[39] In the cobalt-based synaptic transistor, we suggest that the higher channel conductance modulation is related closely to the electrochemical doping of the channel involving proton diffusion. [26, 40–44] During positive gating, the proton intercalates into the channel and raises the channel conductance by introducing external electrons.[45] The proton stays inside the channel even after gating, resulting in a nonvolatile change of channel conductance. In contrast, the proton is extracted from the channel under a negative gating, decreasing the channel conductance and restoring it to the initial value. The schematic illustration of the electrochemical doping process of the channel is shown in the supplementary section, Figure S3.

To study the effect of gate voltage sweeping rate, the transfer curves were measured by sweeping $V_G$ at different rates (0.1, 0.05, 0.01, and 0.003 V/s) as shown in Figure 1c, and the corresponding leakage current was monitored (Figure 1d). The results are summarized in Figure 1e, indicating the area under the transfer curves increases inversely (left), and the leakage current curve increases directly (right) as a function of sweeping rate. It was observed that the slowest (fastest) sweeping rate results in a giant (tiny) transfer curve loop. At a slower sweeping rate, the device is given more time to respond to the changing gate voltage, resulting in more electrochemical doping, hence causing a larger change in channel conductance.

Multilevel, nonvolatile states are crucial for the emulation of synaptic functionalities and analog computation. By applying a series of negative gate voltage pulses of different amplitudes, in a sequence of 0.5, -1, -1.5, -2, and -2.5 V, the multilevel nonvolatile states were realized in the synaptic device, Figure 1b. Each gate pulse has $t_p$ = 60 s and are spaced apart by 120 s. A constant $I_{SD}$ = 160 µA was applied to measure the channel conductance. The channel conductance decreased during negative gating (more decrement at higher amplitude) while retained the same state at zero gating, exhibiting nonvolatile nature. The retention of a separate state is shown in Figure S4, showing excellent retention for at least $3 \times 10^3$ s. The observed multilevel behavior is mainly due to the controlled interfacial electrochemical doping of the channel under programmed gate voltages.[46] The above results demonstrate multiple distinct,

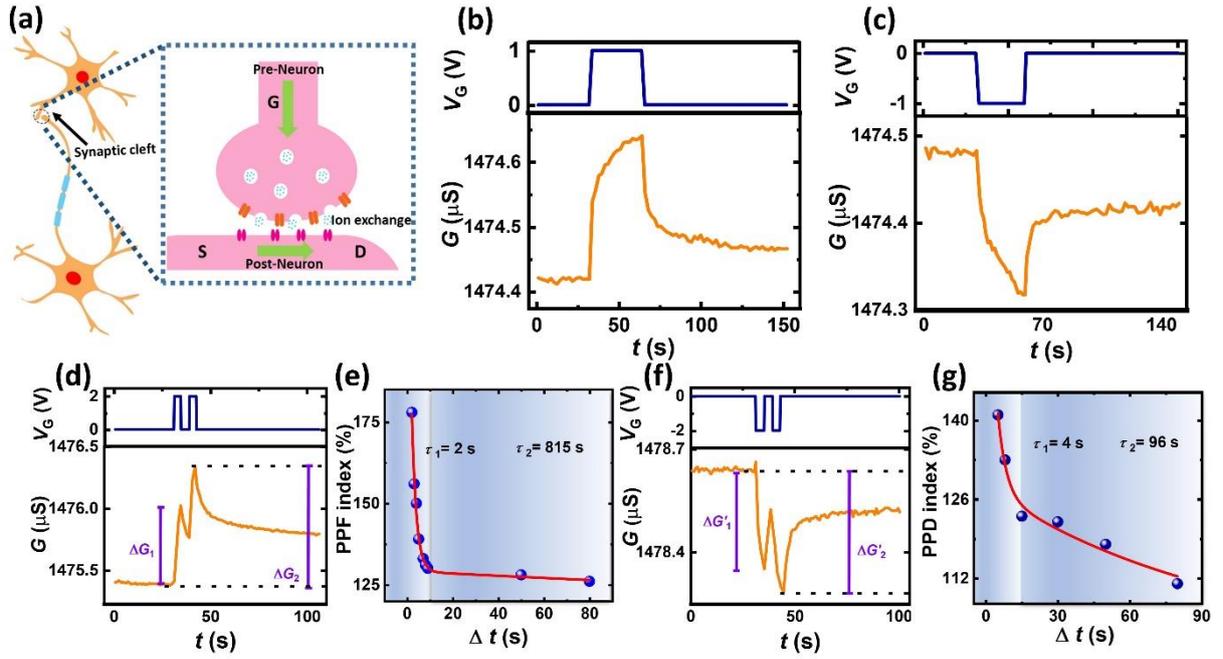

**Figure 2** (a) Schematic illustration of signal transmission across biological synapse, the functional connection between two adjoining neurons. (b) Excitatory post synaptic conductance (EPSC) triggered by a positive gate pulse (1V, 30 s), $I_{SD}$ =160 µA. (c) Inhibitory post synaptic conductance (IPSC) triggered by a negative gate pulse (-1V, 30 s), $I_{SD}$ =160 µA. (d) Paired pulse facilitation (PPF) exhibited by applying a pair of identical positive gate pulses (2 V, 5 s), 2 s apart. $\Delta G_1$ and $\Delta G_2$ are relative modulation evoked by to 1st and 2nd pulse, respectively. (e) PPF index is plotted as a function of time interval ($\Delta t$) between pulses. (f) Paired pulse depression (PPD) behaviour exhibited by applying a pair of identical negative gate pulses (-2 V, 5 s), 2 s apart. $\Delta G'_1$ and $\Delta G'_2$ is relative modulation caused by 1st and 2nd pulse, respectively. (g) PPD index is plotted as a function of time interval ($\Delta t$) between pulses.

nonvolatile states and show the memory is suitable for multilevel memory applications with good retention.

Figure 2a shows the schematic diagram of signal transmission across a synapse, the functional connection between two adjoining neurons. The signal transmission across synapse starts with firing of an action potential on the presynaptic neuron. The presynaptic neuron releases neurotransmitters that diffuse across the synaptic cleft and docks with the receptor of the postsynaptic neuron. This triggers a potential drop across the postsynaptic membrane called postsynaptic potential (PSP). The PSP amplitude depends on the connection strength of the adjoining neurons, termed synaptic weight. Synaptic plasticity is the ability of a synapse to change its synaptic weight in response to external stimuli; it is considered as the basis for learning and memory of the human brain.[47] It is broadly classified into short-term plasticity and long-term plasticity, depending on the retention time of the synaptic weight.

Here in a cobalt-based synaptic transistor, gating controlled channel conductance modulation resembles a change in synaptic weight in the biological synapse. The channel conductance and gate voltage are analogous to synaptic weight and action potential, respectively. The mobile ions play the role of neurotransmitters. The PSP results in a change in the channel conductance termed postsynaptic conductance (PSC).[48] If the postsynaptic conductance is excitatory, it is termed excitatory postsynaptic conductance (EPSC), and if inhibitory, it is termed inhibitory postsynaptic conductance (IPSC). A typical EPSC is demonstrated in Figure 2b by firing a short positive gate pulse (1 V, 5 s) on the gate electrode. A small source-drain current ($I_{SD}$ = 160 µA) was applied to monitor the channel conductance. During gating, the channel conductance was raised to the peak and gradually decayed to the resting state. Analogously, a typical IPSC is demonstrated in Figure 2c by firing a short negative gate pulse (-1 V, 5 s). The channel conductance decreased to the lowest value with gating and gradually recovered to the resting state. Before gating, the anions and cations were distributed randomly in the IL. During gating, the ions get accumulated at the channel/IL interface and provide additional charge carriers in the channel, resulting in channel conductance modulation. However, after gating, the ions diffuse back to the equilibrium state, restoring the channel conductance to the initial value. EPSC and IPSC are complementary functions that underlie successful signal transmission across the synapse.

Short-term plasticity (STP) is defined as strengthening or weakening the synaptic weight for a shorter time, ranging from a few milliseconds to minutes.[49] In the human brain, STP is used in various computations, working memory, and short-term memory. Paired pulse facilitation (PPF) is a vital form of STP. This is well studied for decoding temporal information in auditory and visual signals. [47] PPF depicts a phenomenon where a pair of identical presynaptic pulses are applied in rapid succession; the conductance modulation evoked by the second pulse is higher than the first one. PPF behavior was investigated in our cobalt synaptic transistor by applying a pair of identical pulses (2 V, 10 s) with $\Delta t$ = 2 s, as shown in Figure 2d. It is seen that the conductance modulation evoked by the second pulse ($\Delta G_2$) is higher than the first one ($\Delta G_1$) relative to the base value ($V_G$ = 0). The higher value of $\Delta G_2$ was due to the addition of residual ions (by the first pulse) to the ions accumulated by the second pulse when the second pulse was fired much before the ion relaxation. PPF index is defined as $\Delta G_2/\Delta G_1 \times 100\ \%$ and was measured by varying the time interval ($\Delta t$) between pulses and shown in Figure 2e. We have obtained the highest PPF index (170 %) at the smallest time interval (2 s) attributed to more residual ions. The PPF index variation with $\Delta t$ was fitted to the standard PPF equation,[50] defined as

$$PPF = 100 + C_1 \cdot \exp\left(-\frac{\Delta t}{\tau_1}\right) + C_2 \cdot \exp\left(-\frac{\Delta t}{\tau_2}\right) \qquad (1)$$

Where $C_1$ and $C_2$ are the amplitudes of initial facilitation, and $\tau_1$ and $\tau_2$ are relaxation time constants of different phases. The PPF index undergoes a double exponential decay, corresponding to rapid decay ($\tau_1 = 2$ s) and slow decay ($\tau_2 = 815$ s). In contrast, paired-pulse depression (PPD) was demonstrated in the cobalt synaptic transistor by applying a pair of negative pulses (-2 V, 10 s) with $\Delta t = 2$ s, where $\Delta G_2$ was higher than $\Delta G_1$, Figure 2f. The variation of PPD index with $\Delta t$ was fitted with equation 1 and showed a rapid ($\tau_1 = 4$ s) and slow ($\tau_2 = 96$ s) phase decay (Figure 2g). Such PPF and PPD behavior in cobalt-based synaptic transistors are pretty similar to the short-term plasticity behavior in biological synapses. However, the decay time constants are larger than biological synapses due to the slow diffusion of ions in the device.

The human memory system mainly consists of two forms of memory called short-term memory (STM) and long-term memory (LTM). The schematic diagram of the multistore model of the human memory system is shown in Figure S5. Different memory types are associated with different states of synaptic plasticity: STM is associated with short-term plasticity, while LTM is associated with long-term plasticity.[51] A transition from STP to LTP (and hence from STM to LTM) has been demonstrated in cobalt-based synaptic transistors by variation of gate pulse the amplitude, duration, and number.

Synaptic plasticity can be tuned by varying the gate pulse amplitude. Figure 3a shows EPSC triggered by series of positive gate pulses of the same duration ($t_p = 5$ s) and different amplitudes (0.5, 1.0, 1.5, 2.0, and 2.5 V). $I_{SD} = 160$ µA was applied to monitor the channel conductance. The EPSC peak and retention have increased systematically with increasing gate pulse amplitude. In contrast, IPSC triggered by negative gate pulses of same duration ($t_p = 5$ s) and different amplitudes (-0.5, -1.0, -1.5, -2.0, and -2.5 V) is shown in supplementary (Figure S6a). The conductance modulation ($\Delta G$) evoked by a gate pulse is defined as the difference between the base (at $V_G = 0$) and peak value. The $\Delta G$ increased monotonously with gate pulse amplitude in both gate polarities due to stronger electrochemical doping at higher gate amplitude, demonstrating the transition from STP to LTP (Figure 3b). From Figure 3a, the conductance modulation was measured immediately at peak value and after 100 s of pulse

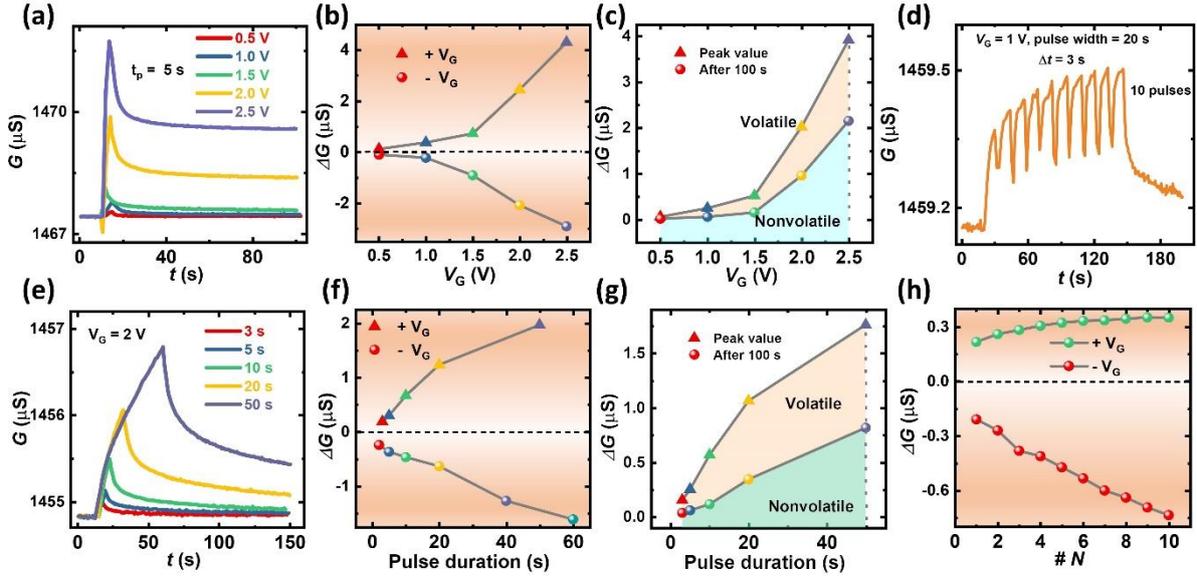

**Figure 3** (a) EPSC triggered by gate pulses of same duration ($t_p$ = 5 s) and different amplitudes (0.5, 1.0, 1.5, 2.0 and 2.5 V), $I_{SD}$ =160 µA. (b) Gate amplitude dependent channel conductance modulation in both positive and negative polarity. (c) Gate amplitude dependent conductance modulation measured immediately (volatile change) and after 100 s (non-volatile change) of pulse removal, demonstrating transition from STM to LTM. (d) EPSC triggered by a series of 10 consecutive gate pulses (1 V, 10 s), $I_{SD}$ =160 µA. (e) EPSC triggered by gate pulses of same amplitude ($V_G$ = 2 V) and different duration (3, 5, 10, 20 and 60 s), $I_{SD}$= 160 µA. (f) Gate pulse duration dependent channel conductance modulation in both positive and negative polarity. (g) Gate pulse duration dependent conductance modulation measured immediately (volatile change) and after 100 s (non-volatile change) of pulse removal, showing transition from STM to LTM. (h) Gate pulse number dependent channel conductance modulation in both positive and negative polarity.

removal, defining a volatile and nonvolatile change, respectively.[26] The volatile change represents STM, whereas the nonvolatile change represents LTM. It is shown in Figure 3c that the nonvolatile part is increasing significantly with increasing gate amplitude, indicating a transition from STM to LTM.

Pulse duration plays a significant role in synaptic plasticity. Figure 3e shows EPSC triggered by gate pulses of the same amplitude ($V_G$ = 2 V) and different durations (3, 5, 10, 20, and 50 s). The EPSC peak and state retention have the highest value for the most extended pulse duration. In contrast, the IPSC triggered by negative gate pulses of different durations is shown in supplementary (Figure S6b). The conductance modulation (Δ$G$) is summarized as a function of pulse duration in Figure 3f, showing linear dependence for both gate polarities. This demonstrates a transition to LTP due to prolonged electrochemical doping for a longer pulse duration. The conductance modulation was measured at peak (volatile change) and after 100 s (nonvolatile change) of pulse removal and plotted as a function of pulse duration (Figure 3g).

It shows that the nonvolatile part is increasing (decreasing) with increasing (decreasing) of the pulse duration, showing a transition from STM to LTM.

Synaptic plasticity can also be tuned by gate pulse number. Here, we have applied 10 consecutive positive gate pulses (1 V, 10 s) spaced apart by 3 s on the gate electrode and monitored the EPSC, as shown in Figure 3d. The channel conductance increased monotonously with each consecutive pulse. The conductance modulation ($\Delta G$) at each pulse was measured with respect to the base value ($V_G = 0$). In contrast, the IPSC was monitored by applying 10 consecutive negative gate pulses (-1 V, 10 s) with $\Delta t = 3$ s, shown in Figure S6c. The $\Delta G$ increased significantly with pulse number, indicating a transition trend from STP to LTP (Figure 3h). This concludes that, at higher pulse amplitude, duration, and the number of pulses, the $\Delta G$ is higher due to more electrochemical proton doping, resulting in a transition from STP to LTP hence from STM to LTM.

Long-term plasticity is widely considered as the basis for learning and memory in the human brain.[47] This is defined as the persistent modification of the synaptic weight that usually lasts from hours to years. [52] Long-term plasticity is emulated in the cobalt-based synaptic transistor via long-term potentiation (LTP) and long-term depression (LTD). LTP was comprehended by applying 30 consecutive positive gate pulses (1 V, 5 s) spaced apart by 2 s on the gate terminal. A small direct current, $I_{SD} = 160$ µA, was applied to measure the channel conductance. The series of positive gate pulses have significantly facilitated the channel conductance. However, after pulse removal, the channel conductance decayed initially and remained at a higher conducting state for a longer time (till we measure), showing its long-term effect (Figure 4a). In contrast, LTD was realized by applying 30 consecutive negative gate pulses (-1 V, 5 s), spaced apart by 2 s resulting in a drastic decrement of channel conductance. Furthermore, the conductance remained at a lower conducting state for a longer time (till we measure) after pulse removal, shown in Figure 4b. In Figure 4c, the channel conductance was decreased by applying 20 consecutive negative gate pulses (-1 V, 5 s) with $\Delta t = 2$ s. Again, 20 consecutive positive gate pulses (1.4 V, 5 s), with $\Delta t = 2$ s, were applied to bring it back to the initial state.

This shows that the cobalt synaptic transistor can imitate essential long-term plasticity functions, and the mechanism can be explained as follows. During LTP, the channel conductance was increased monotonously with subsequent gate pulses due to the intercalation

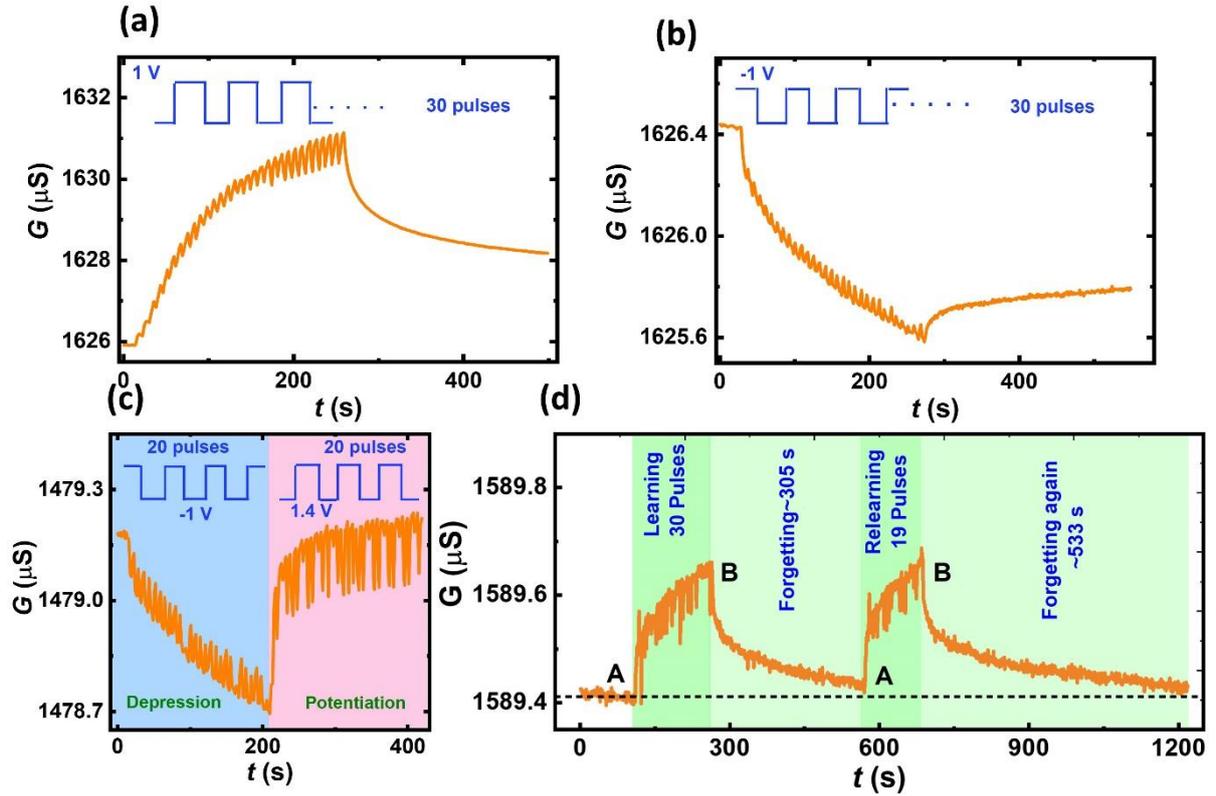

**Figure 4** (a) Long term potentiation (LTP) - Channel conductance versus time stimulated by 30 consecutive positive gate pulses (1V, 5 s) spaced apart by 2 s, $I_{SD}$ =160 µA. (b) Long term depression (LTD)- Channel conductance versus time stimulated by 30 consecutive negative gate pulses (-1V, 5 s) spaced apart by 2 s, $I_{SD}$ =160 µA. (c Demonstrating depression (D) and potentiation (P) of synaptic weight by applying twenty consecutive negative gate pulses (-1 V, 5 s), $\Delta t$ = 2 s for D and twenty consecutive positive pulses (1.4 V, 5 s), $\Delta t$ = 2 s for P. (e) Emulation of psychological behaviour, learning, forgetting, re-learning in cobalt based synaptic transistor.

of more protons into the channel. However, the proton stays inside the channel even after gating, resulting in a long-term change of channel conductance. During LTD, the channel conductance was decreased monotonously by extraction of protons from the channel with subsequent negative pulses.

We have successfully imitated the psychological learning, relearning, and forgetting behavior of the human brain in the cobalt synaptic transistor.[18, 53] It was demonstrated by applying a series of positive gate voltage pulses. As shown in Figure 4d, by applying 30 consecutive positive gate pulses (1 V, 5 s), the channel conductance was increased from a lower conducting state (A) to a higher conducting state (B), considered as learning. After learning, the system spontaneously decays to state A in 305 s, considered as forgetting. Again, 19 consecutive positive gate pulses were applied on the gate electrode to change the channel conductance from state A to state B for relearning. It is noteworthy that only 19 pulses were used to realize relearning compared to 30 pulses for the learning. After relearning, the system decays

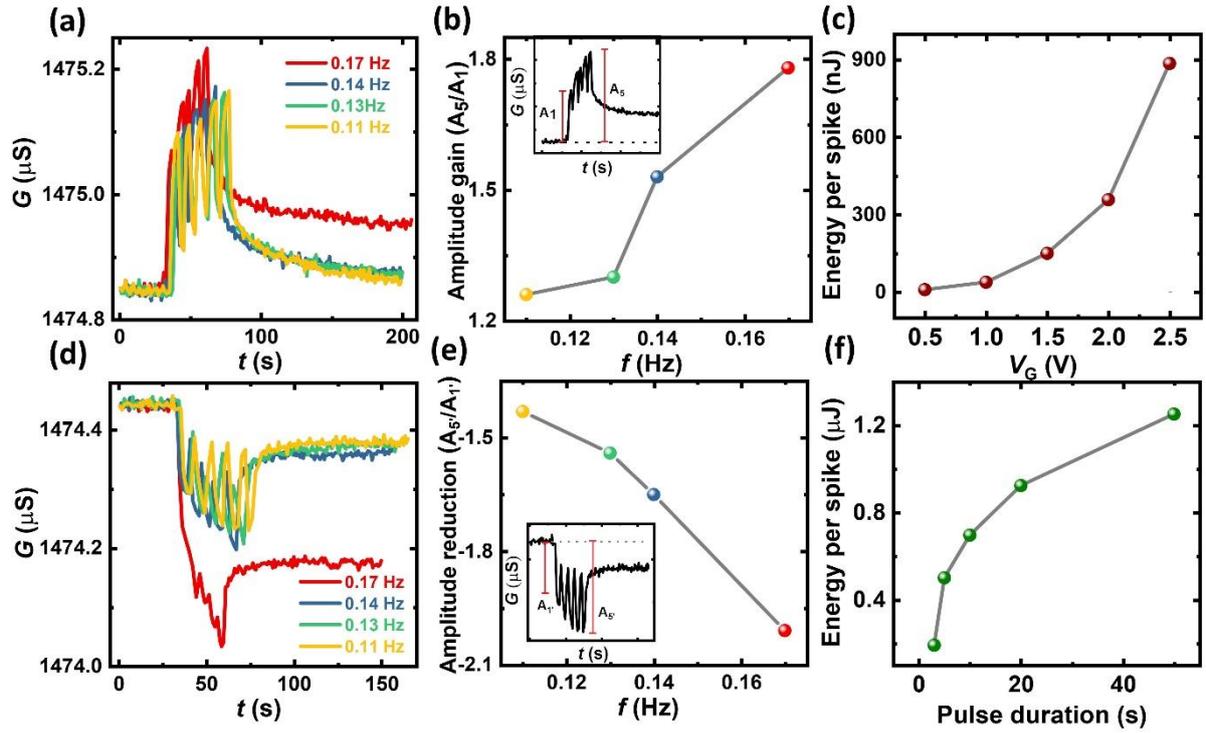

**Figure 5** (a) EPSC response to stimulus train (1 V, 2 s) fired at different frequencies. (b) Amplitude gain ($A_5/A_1$) as a function of frequency, acting as high pass filter. The inset shows $A_1$ and $A_5$ as the conductance modulation evoked by 1st and 5th pulse respectively. (c) Variation of energy consumption per spike as a function of gate pulse amplitude. (d) IPSC response to stimulus trains (- 1 V, 2 s) fired at different frequencies. (e) Amplitude reduction ($A_{5'}/A_{1'}$) as a function of frequency, acting as low pass filter. (f) Variation of energy consumption per spike as a function of gate pulse duration.

spontaneously to state A in 533 s. This implies that for relearning, we need lesser effort and relearning makes the forgetting process slower. Hence, it is easier to remember a forgotten memory than remembering a new memory in the proposed synaptic transistor, similar to the human brain.

The cobalt-based synaptic transistors can perform as a dynamic filter for information transmission. [47] The short-term facilitation/depression phenomenon contributes to high/low pass filtering behavior. The schematic illustration of the filtering behavior of the cobalt synaptic transistor is shown in Figure S7. To demonstrate high pass filtering behavior, the EPSC evoked by positive stimulus trains of different frequencies is shown in Figure 5a. $I_{SD}$ = 160 µA was applied across the source and drain to measure channel conductance. Each stimulus train comprises five gate pulses (1 V, 2 s). EPSC response was increased substantially for higher frequency. The amplitude gain is defined as the ratio $A_5/A_1$, where $A_1$ and $A_5$ are increments of channel conductance relative to the base value (at $V_G$ = 0), after the first and fifth pulse, respectively (Figure 5b inset). As shown in Figure 5b, the amplitude gain increased with increasing frequency, illustrating that the device can easily pass the high-frequency signals. This presents the high pass filtering behavior of the proposed transistor for sophisticated information processing.

On the other hand, Figure 5d shows the IPSC evoked by negative stimulus trains (-1 V, 2 s) of different frequencies to demonstrate low pass filtering behavior. The IPSC response was increased (downward) significantly for a higher frequency. The amplitude reduction is defined as the ratio $A_{5'}/A_{1'}$, where $A_{1'}$ and $A_{5'}$ are channel conductance decrements after the first and fifth pulse, respectively (Figure 5e inset). The amplitude reduction increases (downward) with frequency, emulating the low pass filtering behavior (Figure 5e). All the stimulus train consists of pulses of same voltage and width, whereas the spacing between them is varied. Due to lower spacing between the pulses at higher frequencies, the ions do not get enough time for relaxation, leading to more conductance modulation. Hence, both high and low pass filtering behavior was successfully mimicked in the same device by changing the gate polarity, which is crucial for selective communication.[17, 54, 55]

Energy consumption is one of the most critical performance metrics of a synaptic device. In a three-terminal synaptic transistor, the energy per spike is obtained using the formula $dE = V_G \times I_G \times dt$ and integrating over time, where $V_G$, $I_G$, and $dt$ represent the gate voltage, leakage current, and pulse duration, respectively. [56][57, 58] The energy per spike was calculated for pulses of different amplitudes, showing a rapid decrement with lowering the gate pulse amplitude, as presented in Figure 5c. The energy consumption calculated for pulses of different duration showed a lower value for a shorter duration (Figure 5f). It shows that the energy consumption of the device can be reduced by using gate pulses of lower amplitude and shorter duration. In our cobalt-based synaptic transistor, we have obtained the lowest energy consumption of 6 nJ for a single pulse (0.8 V, 2 s), as shown in Figure S8. However, the energy consumption in our device is large compared to the biological synapse and can be reduced by considering the following factors. For the present study, we have used a huge channel dimension ($\approx$ 1000 µm) and wider gate pulses ($\approx$ seconds). By reducing the device dimension to a submicron scale and using millisecond range gate pulses, the energy consumption of the device can be reduced by several orders of magnitude.

**Conclusions:**

In summary, we have demonstrated a three-terminal synaptic transistor based on a metallic channel such as cobalt for the first time. We have obtained a giant transfer curve loop at the slowest sweeping rate of gate voltage, showing a nonvolatile and reversible change of channel conductance, suitable for emulation of synaptic properties. Multilevel, nonvolatile conducting states have been realized in the device by applying series of gate pulses of different amplitudes. We have emulated several important synaptic functions including EPSC/IPSC, PPF/D, LTP/D, in the proposed synaptic device. Furthermore, a transition from STM to LTM has been demonstrated by varying the gate pulse amplitude and duration. We have successfully mimicked the psychological behavior, learning, forgetting, and relearning of the human brain. Both high and low pass filters have been realized in the same device by changing the gate voltage polarity. The proposed metallic cobalt thin film-based synaptic transistor has great potential in synaptic electronics and neuromorphic computing applications.

**Experimental Details:**

*Device Fabrication:*

The channel having a dimension of 1000 µm *(L)* × 200 µm (*W*) was fabricated using standard photolithography and subsequent lift-off process. The Ta (3 nm)/Co (10 nm)/Ta (2 nm) multilayer stack was deposited on a thermally oxidized silicon substrate using DC magnetron sputtering. Ar pressure of 2 mTorr and DC power of 50 W was used during the deposition process. The three-terminal coplanar (source, drain, and gate electrode lie on the same plane) electrode was defined using photolithography. The electrode consisting of Ta (5 nm)/Cu (90 nm)/Ta (5 nm) stack was deposited using DC magnetron sputtering. After device fabrication, a small drop of as received EMIM-TFSI (Sigma Aldrich) IL was dropped on the device, covering the channel and gate electrode. This makes the device ready for further measurements.

*Electrical Measurements:*

All the electrical measurements were performed at room temperature using Lakeshore Janis ST300 compact cryostats. A vacuum of $10^{-3}$ mbar was maintained to protect the IL from ambient and to reduce noise during data collection. Yokogawa GS20 DC voltage/current source and Keithley 2000 multimeter were used to supply $I_{SD}$ and measure channel conductance, respectively. Meanwhile, Keithley 6517B electrometer was used to apply gate voltage pulses.

The frequency-dependent capacitance measurement of the IL was carried out using GWINSTEK LCR-8105G.

**Acknowledgments:**

The authors acknowledge the support from the CRP grant NRF-CRP21-2018-003 of the National Research Foundation (NRF), Singapore. P.M. thank the Ministry of Education (MoE), India, and the Pratiksha Trust, for the financial support.

**Conflict of Interest**

The authors declare no conflict of interest.